\documentclass[12pt]{iopart}

\usepackage{graphicx}% Include figure files
\usepackage{latexsym}
\usepackage{dcolumn}% Align table columns on decimal point
\usepackage{bm}% bold math

\usepackage{epsfig}

\begin{document}

\title[Effects of $J$-gate potential and interfaces on donor exchange coupling ] { Effects of $J$-gate potential and interfaces on donor exchange coupling in the Kane quantum computer architecture}

\author{L M Kettle\dag\ddag\footnote[3]{To whom correspondence should be addressed (s335321@student.uq.edu.au)}, H-S Goan$\|$, Sean C Smith\ddag, L C L Hollenberg\P and C J  Wellard\P}

\address{\dag Centre for Quantum Computer Technology, University of Queensland, Brisbane QLD 4072 Australia}

\address{\ddag Centre for Computational Molecular Science, University of Queensland, Brisbane QLD 4072 Australia}

\address{$\|$ Centre for Quantum Computer Technology, University of New South Wales, Sydney NSW 2052 Australia}

\address{\P Centre for Quantum Computer Technology, University of Melbourne, Melbourne VIC 3010 Australia}

\begin{abstract}
We calculate the electron exchange coupling for a phosphorus donor pair in Si perturbed by a $J$-gate potential and the boundary effects of the silicon host geometry. In addition to the electron - electron exchange interaction we also calculate the contact hyperfine interaction between the donor nucleus and electron as a function of the varying experimental conditions. Donor separation, depth of the  P nuclei below the silicon oxide layer and $J$-gate voltage become decisive factors in determining the strength of both the exchange coupling and hyperfine interaction - both crucial components for qubit operations in the Kane quantum computer. These calculations were performed using an anisotropic effective-mass Hamiltonian approach. The behavior of the donor exchange coupling as a function of the parameters varied in this work provides relevant information for the experimental design of these devices. 
\end{abstract}

\pacs{03.67.Lx, 71.55.Cn, 85.30.De}

\submitto{\JPCM}
\maketitle

\section{\label{sec:one} Introduction}

Kane's proposal\cite{kane}  of a donor based solid state quantum computer in silicon  has sparked a concerted effort to re-evaluate an atomistic view of impurities in doped silicon electronic devices. In the Kane quantum computer the phosphorus donor nuclear spins act as qubits, and single qubit operations are performed by applying radio frequency magnetic fields resonant with nuclear spin transitions. Two qubit operations are mediated through the electron exchange interaction. Application of voltages to metal gates above the spins ($A$-gates) and between adjacent spins ($J$-gates) perturb the donor electron density around the nucleus, and thus the hyperfine and exchange interactions can be tuned with an externally applied electric field.

We study these two gate-controlled interactions crucial for qubit operations: the hyperfine interaction between P nuclear spin and donor electron spin, and the exchange interaction between adjacent donor electrons. Here we modelled the effect of either an $A$ or $J$-gate voltage as well as the effect of the location of the qubit in the silicon wafer device to determine the sensitivity of these interactions in relation to these parameters.

In \sref{sec:two} we discuss the approach we took to obtain the phosphorous donor ground state in the silicon wafer device. The donor wave function was expanded in a basis of deformed hydrogenic orbitals following Faulkner's approach\cite{faulkner} using an anisotropic effective mass Hamiltonian. To include the effect of the electric field and interface regions into the Hamiltonian we modeled the application of an electrostatic potential to the metallic gates above the qubits using TCAD,\cite{tcad} and used a step potential to model the $\mbox{Si}/\mbox{SiO}_2$ and Si/back gate barrier. 

\Sref{sec:three} discusses how we calculated the contact hyperfine interaction and the exchange interaction for the donor pair. Here we performed a Heitler-London calculation of the exchange coupling with the application of a $J$-gate potential, to study how the electrostatic potential enhances the exchange coupling.  

We present the numerical results for the hyperfine and exchange interaction in \sref{sec:five} and \ref{sec:fivea}. We explore how the application of a gate voltage and the qubit position affects these two interactions. We can examine the selectivity of the gate potential by comparing the hyperfine interaction with the application of either an $A$ or $J$-gate voltage.\cite{me, lloyd} We can also examine the connectivity between the donor pair by calculating the exchange splitting at varying  $J$-gate voltage, inter-donor separation and donor depth. Finally we summarise our major findings in \sref{sec:six}.
 
There is a great deal of attention on modelling these interactions, Kane\cite{kane2} makes a qualitative calculation using a hydrogenic approximation for the exchange coupling in bulk Si, without considering the electric field potential. Koiller \emph{et al.}\cite{koiller,koiller2} studied the exchange coupling between a donor pair also in bulk Si, and the absence of an electric field. In their calculations they used an effective mass theory in which the expansion of the ground state donor electron wave function includes the Bloch states of the six conduction band minima. They approximated the coefficients of the Bloch functions using an anisotropic Kohn-Luttinger variational form for the envelope wave function. Wellard \emph{et al.}\cite{cam} have extended these calculations to remove some of their approximations. They obtained the donor electron wave function and bare exchange coupling at zero $J$-gate bias, in order to study the fast exchange oscillations with respect to fabrication strategies. 

Fang \emph{et al.}\cite{fang} calculated the donor electron wave function using the spherical effective mass approximation. They modeled the $J$-gate potential qualitatively as a 1-D parabolic well with its minimum located in the middle of the two donor sites, but did not consider the boundary effects of the silicon host geometry in their calculation. In their work they used an unrestricted Hartree-Fock method with a generalised valence bond wave function to study the two-electron system and calculate the exchange coupling. Parisoli \emph{et al.}\cite{fran} have calculated the effect of the $J$-gate potential, interface regions and donor separation using a spherical effective mass Hamiltonian. We extended this work to include the anisotropy of the effective masses in Si into the Hamiltonian.

We study the effect of application of a 3-D electrostatic potential to the metallic gates above the qubits, and boundary effects of the silicon oxide layer and back gate on the donor electron wave function. We included the anisotropy of the effective masses, the P impurity potential, electric field and interface potentials into the Hamiltonian. We calculated the contact hyperfine interaction and exchange coupling for varying qubit separation, qubit depth and gate voltage. We aim to provide relevant information for experimental engineering of these devices and highlight the significance of environmental factors other than the gate potential which may perturb the donor electron wave function.

\section{\label{sec:two}Faulkner's Method with the applied electric field and silicon host potential}

Using the method outlined previously,\cite{me} we use an anisotropic effective mass Hamiltonian, $H_0$, for the donor in bulk Si and zero field:
\begin{eqnarray}
-\bigg{\lbrack} \frac{\partial^2}{\partial x^2} + \frac{\partial^2}{\partial y^2} + \gamma \frac{\partial^2}{\partial z^2} + \frac{2}{r} \bigg{\rbrack} \Psi(\mathbf{r}) &=& E \Psi(\mathbf{r}), \nonumber \\ && \label{a1}
\end{eqnarray}
where $\epsilon = 11.4$ is the dielectric constant, and $\gamma = m_{\bot} / m_{\parallel} = 0.2079$. Here we are using atomic units, where the unit of length, $a_B  =  \hbar^2 \epsilon / m_\perp e^2= 31.7 ${\AA} and the unit of energy, $E_B = m_{\perp} e^4 / 2 \hbar^{2} \epsilon^2  = 19.94$meV.  

Following Faulkner's approach\cite{faulkner} we expanded the donor electron wave function in a basis of 91 deformed hydrogenic orbitals:
\begin{eqnarray}
\Psi(\mathbf{r})& =& \left( \frac{\beta}{\gamma}\right)^{1/4} \sum_{n,l,m} C_{nlm} \psi_{nlm}(x,y, \sqrt{ \frac{\beta}{\gamma} } z,a),
\end{eqnarray}
where $\psi_{nlm}(x,y,z,a) = R_{nl}(a,r) Y_{lm}(\theta, \phi)$, are the normalised hydrogenic orbitals, $C_{nlm}$ are the expansion coefficients for our basis,  $a$ is the effective Bohr radius in the $x$, $y$ directions, and $\beta$ is an adjustable parameter which gives the effective Bohr radius $b$ in the $z$ direction. \Eref{a1} was solved variationally using a basis of 91 deformed hydrogenic orbitals for the donor in zero field, to give a ground state energy $E = -31.23\mbox{meV}$, and effective Bohr radii: $ a = 23.81${\AA}  and $ b =13.68 ${\AA}.\cite{me}

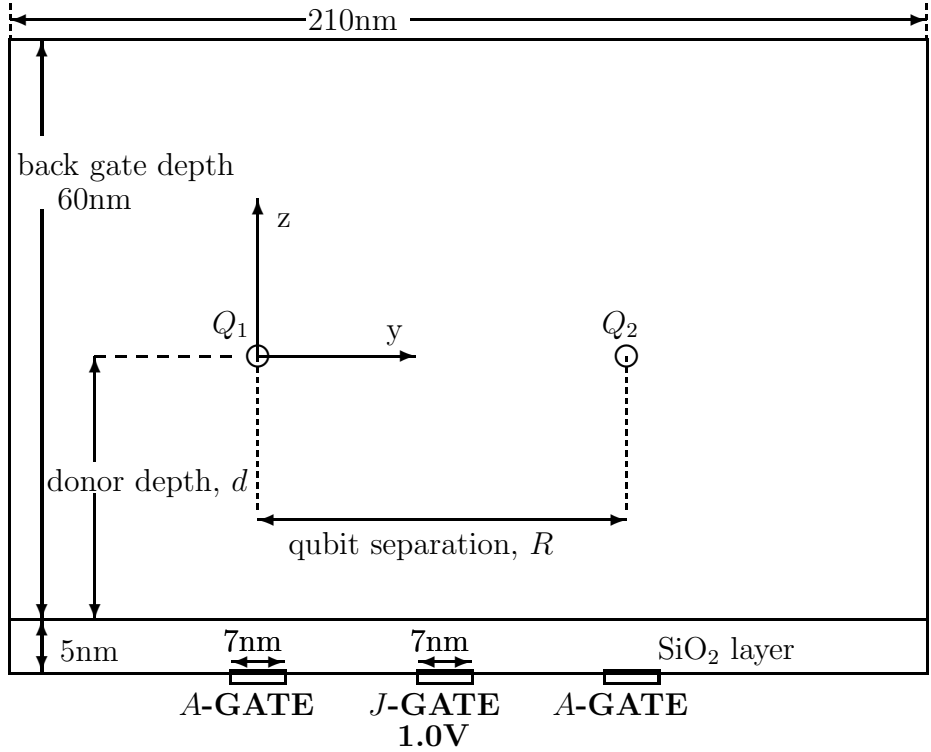
\begin{figure} [t!]
\begin{center}

\setlength{\unitlength}{0.14cm}
\begin{picture}(100,70)
\thicklines
\put(6,7){ \framebox(87,60) {}}

\put(7,12){\line(1,0){87}}

\multiput(7,67)(0,1){4}{\line(0,1){0.5}}
\multiput(94,67)(0,1){4}{\line(0,1){0.5}}

\put(35,69){\vector(-1,0){28}}
\put(45,69){\vector(1,0){49}}
\put(40,69){\makebox(0,0) {210nm   }}

\put(30.5,37){\vector(1,0){15}}

\put(30.5,37){\vector(0,1) {15}}

\put(33.5,50){\makebox(0,0) {z   }}

\put(43.5,39){\makebox(0,0){y}}

\put(10,7) {\vector(0,1){5}}
\put(10,8) {\vector(0,-1){1}}

\put(75,9) {\makebox(0,0){ $\mbox{SiO}_2$ layer }}
\put(15,9) {\makebox(0,0){5nm }}

\put(10,58) {\vector(0,1){9}}
\put(10,51) {\vector(0,-1){39}}
\put(18,55) {\makebox(0,0){ back gate depth }}
\put(15.5,52) {\makebox(0,0){60nm }}

\put(30.5,37){\circle{2}}
\put(65.5,37){\circle{2}}

\put(35.5,21.5){\vector(-1,0) {5}  }
\put(30.5,21.5){\vector(1,0) {35}  }
\put(46.,19.){\makebox(0,0){qubit separation, $R$}}

\multiput(30.5,37)(0,-1){15}{\line(0,-1){0.5}}
\multiput(65.5,37)(0,-1){15}{\line(0,-1){0.5}}

\put(28.5,40){\makebox(0,0){\bfseries {$Q_1$}  }}
\put(65.5,40){\makebox(0,0){\bfseries {$Q_2$}  }}

\put(15,27) {\vector(0,1){10}}
\put(15,24) {\vector(0,-1){12}}

\multiput(15,37)(2,0){7}{\line(1,0){1}}

\put(20,25) {\makebox(0,0){donor depth, $d$}}

\multiput(28,6)(17.75,0){3}{\framebox(5,1){}       }

\multiput(28,8.)(17.75,0){2}{\vector(1,0){5}}
\multiput(29,8.)(17.75,0){2}{\vector(-1,0){1}}

\multiput(30,10)(17.75,0){2}{\makebox(0,0){7nm}}
\multiput(30,10)(17.75,0){2}{\makebox(0,0){7nm}}

\multiput(29.5,4)(35.5,0){2}{\makebox(0,0){ \bfseries {$A$-GATE}  }            }

\put(47.75,1){\makebox(0,0){\bfseries{1.0V} }}

\put(47.25,4) {\makebox(0,0){\bfseries{ $J$-GATE} }}

\end{picture}
\end{center}

\caption{\label{fig:fige1} Schematic design parameters implemented in TCAD to model the Kane computer architecture.}

\end{figure}

To accommodate the effect of the applied field and the boundaries on the donor electron wave function it is necessary to use more than one simple bulk ground state wave function to describe the envelope function. The method we used is advantageous because we expand the envelope wave function in a basis of deformed hydrogenic orbitals which have the flexibility to distort with the applied fields. We use the zero field effective Bohr radii and diagonalise the single donor electron Hamiltonian including the electric field. 

To include the effect of an electric field and the silicon host, we constructed an additional Hamiltonian matrix, $H_1$, with its elements given by:
\begin{eqnarray}
\lefteqn{ \langle n'l'm' \vert H_1 \vert nlm   \rangle } \nonumber \\
&=& \sqrt{ \frac{\beta}{\gamma} }\int dx^3 \psi^*_{n'l'm'}(x,y, \sqrt{ \frac{\beta}{\gamma} } z,a) V_{elec}(y,z) \nonumber \\
&& \times \psi_{nlm}(x,y, \sqrt{ \frac{\beta}{\gamma} } z,a),  \label{ef1}
\end{eqnarray}
where  $V_{elec}(y,z)$ is the electric field potential generated from TCAD, and here we also add an additional term to model the  $\mbox{SiO}_2$ layer and the back gate as a step  function with height 3.25eV.\cite{me,lloyd,Yu} The lateral edges of the silicon lattice were assumed to extend infinitely in the $y$-direction, and the potential in 2-D from TCAD is assumed to have a ``thickness'' in the third dimension ($x$) of 1$\mu$m. \Fref{fig:fige1} shows the 2-D device scheme implemented in TCAD used to model the application of voltages to the $A$ or $J$-gate above qubit, $Q_1$, the metallic gates were modelled as thin wires in the $x$-direction.

The new Hamiltonian $H = H_0+H_1$ was then diagonalised to find the perturbed single donor electron ground state for each particular gate voltage and qubit position. The location of the interfaces in \fref{fig:fige1} splits the degeneracy of the two conduction band minima along the $z$-axis relative to the other four along the $x$ and $y$-axis, in the lower $A_1$ ground state in zero electric field.\cite{koiller} We expect that at the shallow donor depths we consider, the greatest restriction on the donor electron will be the interface regions. Here we expand around the conduction band minimum in the $z$-direction from $Q_1$ to the silicon oxide layer. This was done so that the smaller effective Bohr radius, $b$, would be in the direction toward the silicon oxide and back gate barrier. Using this convention the donor wave function is lower in energy since there is less overlap of the wave function into the interface regions. 

\section{\label{sec:three} Calculation of the hyperfine interaction coupling and exchange splitting }

For the Si:P quantum computer to be feasible, quantum operations have to be able to be applied selectively to particular nuclear spins, and connectivity between nuclear spins via electron-mediated coupling must be established. To achieve both these goals it is necessary to study the degree of selectivity and connectivity that can be controlled by applying electric fields to metal gates above ($A$-gates) and adjacent ($J$-gates) to spins. Furthermore, it is shown in this paper that the qubit location in the device in relation to each other (inter donor separation) and to the gates (donor depth below the silicon oxide barrier), also has a significant influence on the donor electron wave function.

\subsection{Calculation of the contact hyperfine interaction}

Since we use effective mass theory, instead of calculating the contact hyperfine coupling, $A(V)$, directly we calculate the relative shift in $A(V)$ with the potential applied and assume this shift will be similar to those of the true wave function.\cite{lloyd,me} Thus we need to calculate:
\begin{eqnarray}
A(V) &=& \frac{|\Psi(V,0)| ^2}{|\Psi(0,0)|^2 }A(0), \label{j1} 
\end{eqnarray}
where $A(0)/h = 28.76$MHz is determined for $^{31}\mbox{P}$ in silicon from experimental data,\cite{larionov,kane} and $\Psi(V,r)$ are the donor envelope wave functions calculated by our method.

The contact hyperfine interaction was calculated for the varying $J$-gate voltage, inter donor separation $R$, and donor depth below the silicon oxide layer, and compared with our previous results for similar calculations at varying $A$-gate voltages.\cite{me}

\subsection{Calculation of the exchange splitting for an impurity pair}

In this section we employ a Heitler-London (H-L) treatment of the two electron donor pair wave function, using the two single donor ground state wave functions perturbed by the electric field as our basis. Since the donor ions are generally well separated in the silicon wafer device we can justify using H-L theory to describe the two electron system as the symmetrised and anti-symmetrised products of the single donor orbitals at each qubit ($\Psi^{Q_1}(\mathbf{r})$ and $\Psi^{Q_2}(\mathbf{r})$) calculated with the electric field applied. The singlet and triplet impurity donor pair wave functions are given by:\cite{slater}
\begin{eqnarray}
&& \Psi(r) = \Psi^{orbit}_{^S_T} \chi^{spin}_{^S_T},  \nonumber \\
\mbox{where} && \nonumber \\
\Psi^{orbit}_{^S_T} &=& \frac{1}{\sqrt{2 (1 \pm S^2) } } \bigg{\lbrack} \Psi^{Q_1}(\mathbf{r_1}) \Psi^{Q_2}(\mathbf{r_2 - R}) \pm  \Psi^{Q_1}(\mathbf{r_2}) \Psi^{Q_2}(\mathbf{r_1 - R}) \bigg{\rbrack},  \\ 
 S &= &\int \Psi^{Q_1}(\mathbf{r}) \Psi^{*Q_2}(\mathbf{r - R}) dr^3. \nonumber 
\end {eqnarray}
Here $\Psi^{Q_1}(\mathbf{r},V)$ and $\Psi^{Q_2}(\mathbf{r},V)$ are the single wave functions calculated using our basis of deformed hydrogenic orbitals, and diagonalising the Hamiltonian for the varying voltages at the $J$-gate and qubit position. We observe that $\Psi^{Q_1}(x,y,z) = \Psi^{Q_2}(x,-y,z)$, as the donor wave functions on adjacent nuclei are mirror images about the $y$-axis when a voltage is applied to the  $J$-gate (see \fref{fig:fige1}).

To calculate the exchange splitting between the ground singlet and triplet states for an impurity pair of donors in silicon we use the H-L formula:\cite{slater}
\begin{eqnarray}
J(R) &=& E_T - E_S  \nonumber \\
 &=&   \langle \Psi_T | H_{2e} | \Psi_T \rangle - \langle \Psi_S | H_{2e} | \Psi_S \rangle   \nonumber \\
&=& \frac{2}{1-S^4 } \left( S^2 K_0 - K_1 \right), \label{j2} \\
 \mbox{where:}  && \nonumber \\
H_{2e} &=& -\nabla_{anis}^2(\mathbf{r_1}) -\nabla_{anis}^2(\mathbf{r_2}) - \frac{2}{|\mathbf{r_1}|} - \frac{2}{|\mathbf{r_2}|} - \frac{2}{|\mathbf{r_1 - R}|}- \frac{2}{|\mathbf{r_2 - R}|} \nonumber \\
 & + & \frac{2}{|\mathbf{r_1 - r_2}|} + V_{elec}(\mathbf{r_1}) + V_{elec}(\mathbf{r_2}), \nonumber \\
K_0 &=& \int | \Psi^{Q_1}(\mathbf{r_1}) |^2 |\Psi^{Q_2}(\mathbf{r_2 - R})|^2 \Theta dr_1^3 dr_2^3,   \nonumber \\
K_1 & = &  \int \Psi^{*Q_1}(\mathbf{r_2}) \Psi^{*Q_2}(\mathbf{r_1 - R}) \Psi^{Q_1}(\mathbf{r_1}) \Psi^{Q_2}(\mathbf{r_2 - R})  \Theta dr_1^3 dr_2^3 , \nonumber \\
\Theta &=& \frac{2}{ |\mathbf{r_2-r_1}|} - \frac{2}{ |\mathbf{r_1-R}|} - \frac{2}{|\mathbf{r_2}|} . \nonumber
\end {eqnarray}

\section{\label{sec:five} Results obtained varying gate voltage and inter donor separation}

To demonstrate the effect of $J$-gate voltage and inter donor separation on the donor electron ground state we calculated the perturbed single electron donor ground states as a function of these external factors. The calculations in this section were obtained at a donor depth of 20nm. Once the perturbed ground state under the applied field was obtained we calculated the contact hyperfine interaction and the exchange splitting (using \eref{j1} and \eref{j2} respectively), for the impurity donor pair, to optimise and determine the experimental conditions needed to control the nuclear spins coupling to the donor electron spin, via the hyperfine interaction, and to other nuclei via the electron-mediated exchange interaction.

So far we have only considered the effect of the $A$ or $J$-gate independently. The smaller inter donor distances ($R \leq 14$nm) are only possible if the gate dimensions can be reduced to prevent overlapping gates. For this work we have only considered the impact of the $J$-gate without reference to the $A$-gate. In this initial study we aim to give insight into, and identify the relevant factors that contribute to the hyperfine and exchange coupling, which need to be studied more in depth. 

\subsection{Results for the contact hyperfine interaction}

\begin{figure} 
\begin{center}
 \includegraphics[height=3in,width=2.5in,angle=-90]{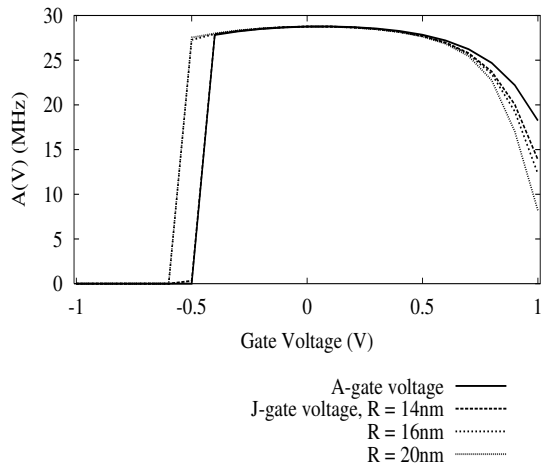} % ,width=7cm,height=11cm,clip=,angle=270}
\end{center}
 \caption{\label{fig:figr3}   Contact hyperfine interaction at varying gate voltage and inter donor separation, for a donor depth of 20nm. }
 \end{figure}

\Fref{fig:figr3} shows the magnitude of the contact hyperfine interaction calculated for varying qubit separation and gate voltage.  As the donor electron density decreases at the P nucleus so does the contact hyperfine interaction. These results reflect the trend that as the qubit moves away from the  $J$-gate, the donor electron wave function has more freedom to move towards the  $J$-gate and distort greater. 

\begin{figure} [t!]
\begin{center}
 \includegraphics[height=3in,width=2.5in,angle=-90]{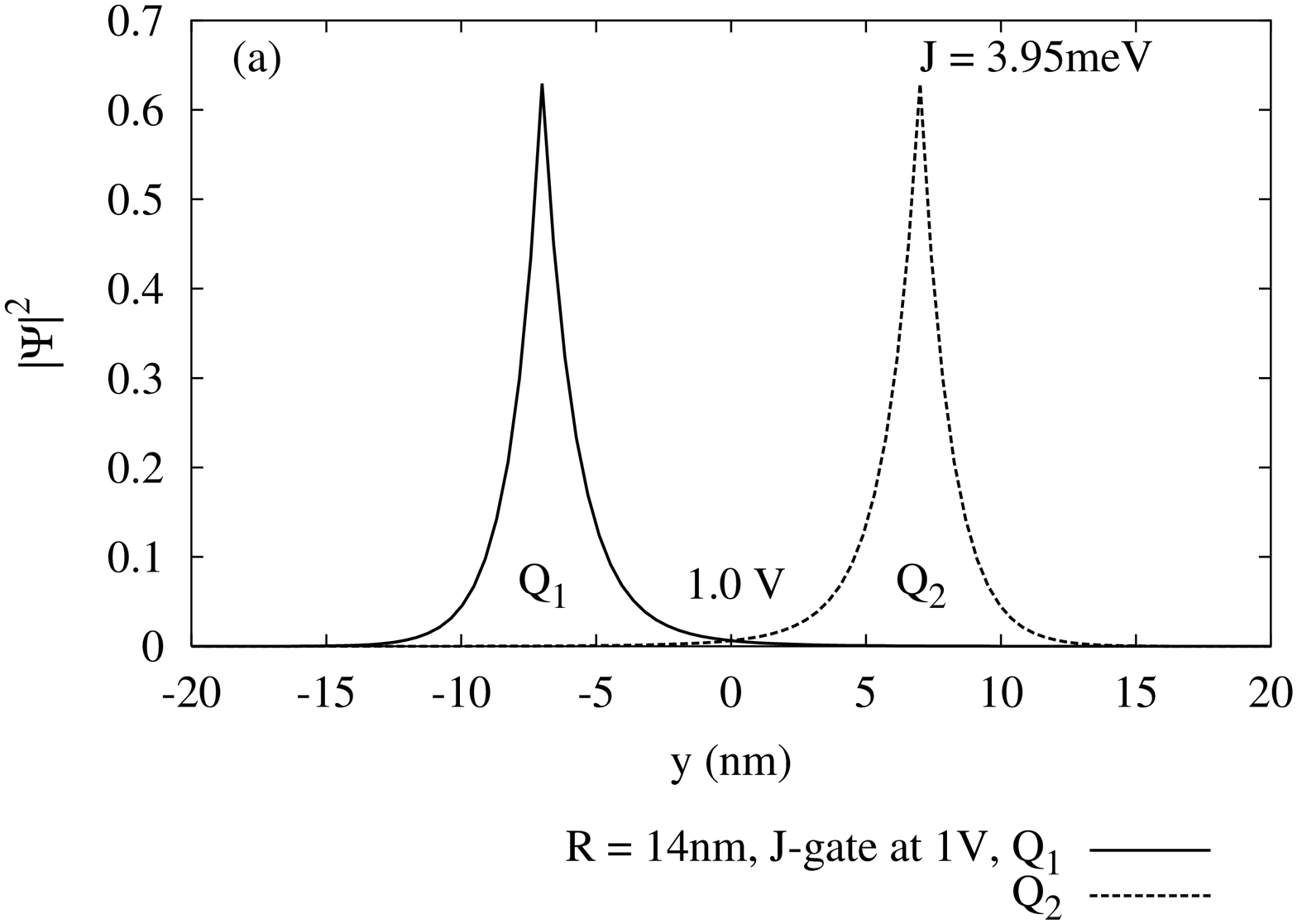} % ,width=7cm,height=11cm,clip=,angle=270}
 \includegraphics[height=3in,width=2.5in,angle=-90]{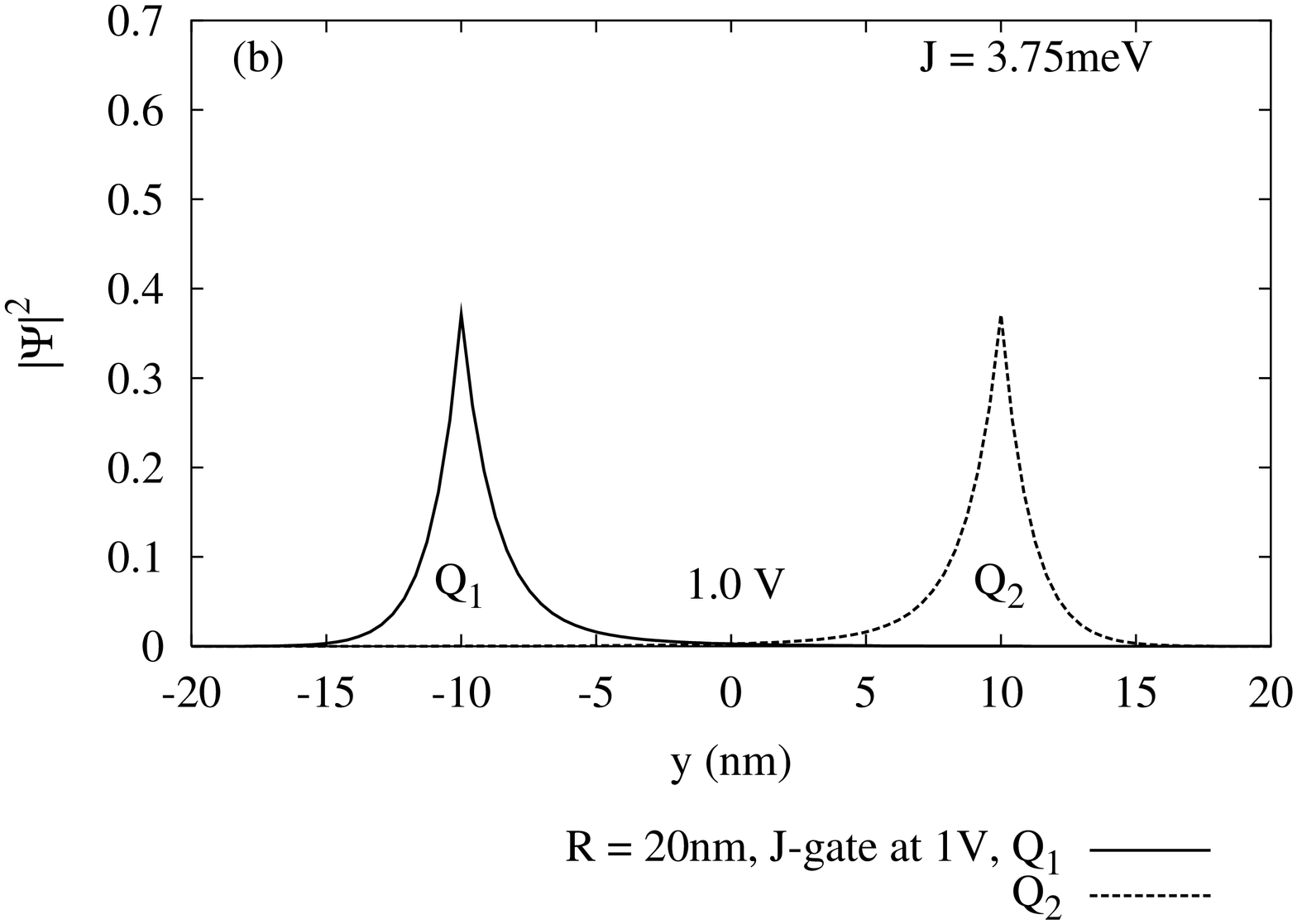} % ,width=7cm,height=11cm,clip=,angle=270}
\end{center}
 \caption{\label{fig:figx4} Ground state electron densities of $Q_1$ and $Q_2$, in $y$-direction at $z=0,x=0$ for an inter donor separation of $R = 14$ and 20nm, a donor depth of $d = 20$nm and a voltage of 1.0 V at the $J$-gate. }
\end{figure}

\Fref{fig:figx4} shows an example of the donor ground state wave functions of $Q_1$ and $Q_2$ for an applied voltage of 1.0 V at the $J$-gate, for two inter donor separations, $R = 14$ and 20nm. From the relative magnitudes of the ground state electron densities of the two qubits we can see that the ground state wave functions for $R = 20$nm have perturbed more towards the $J$-gate voltage. This figure demonstrates how the electron density is perturbed greater for larger inter donor separations, (which also implies a greater distance of the qubits from the $J$-gate) at large positive gate voltages. 

\begin{figure} [t!]
\begin{center}
 \includegraphics[height=3in,width=2.5in,angle=-90]{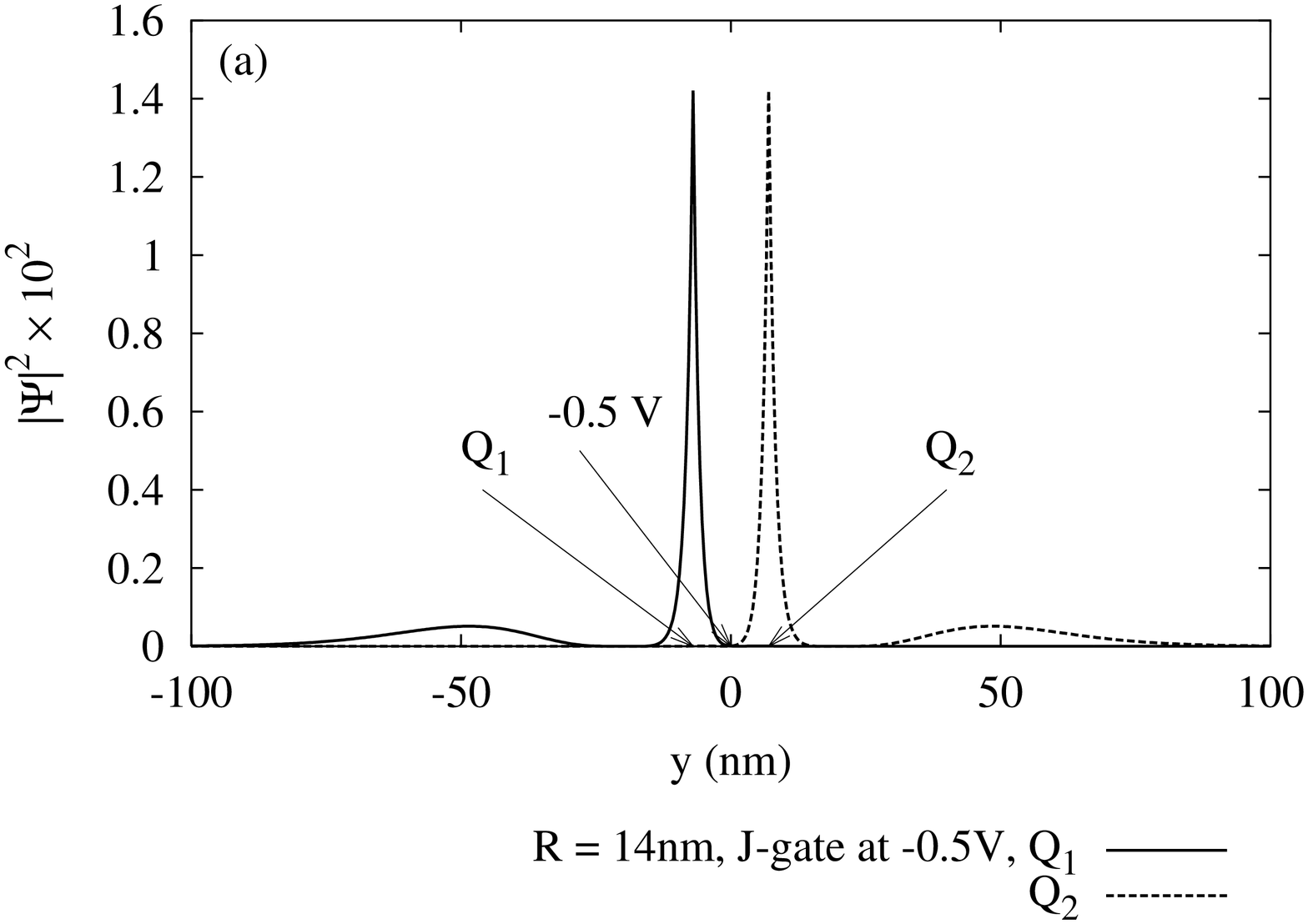} % ,width=7cm,height=11cm,clip=,angle=270}
 \includegraphics[height=3in,width=2.5in,angle=-90]{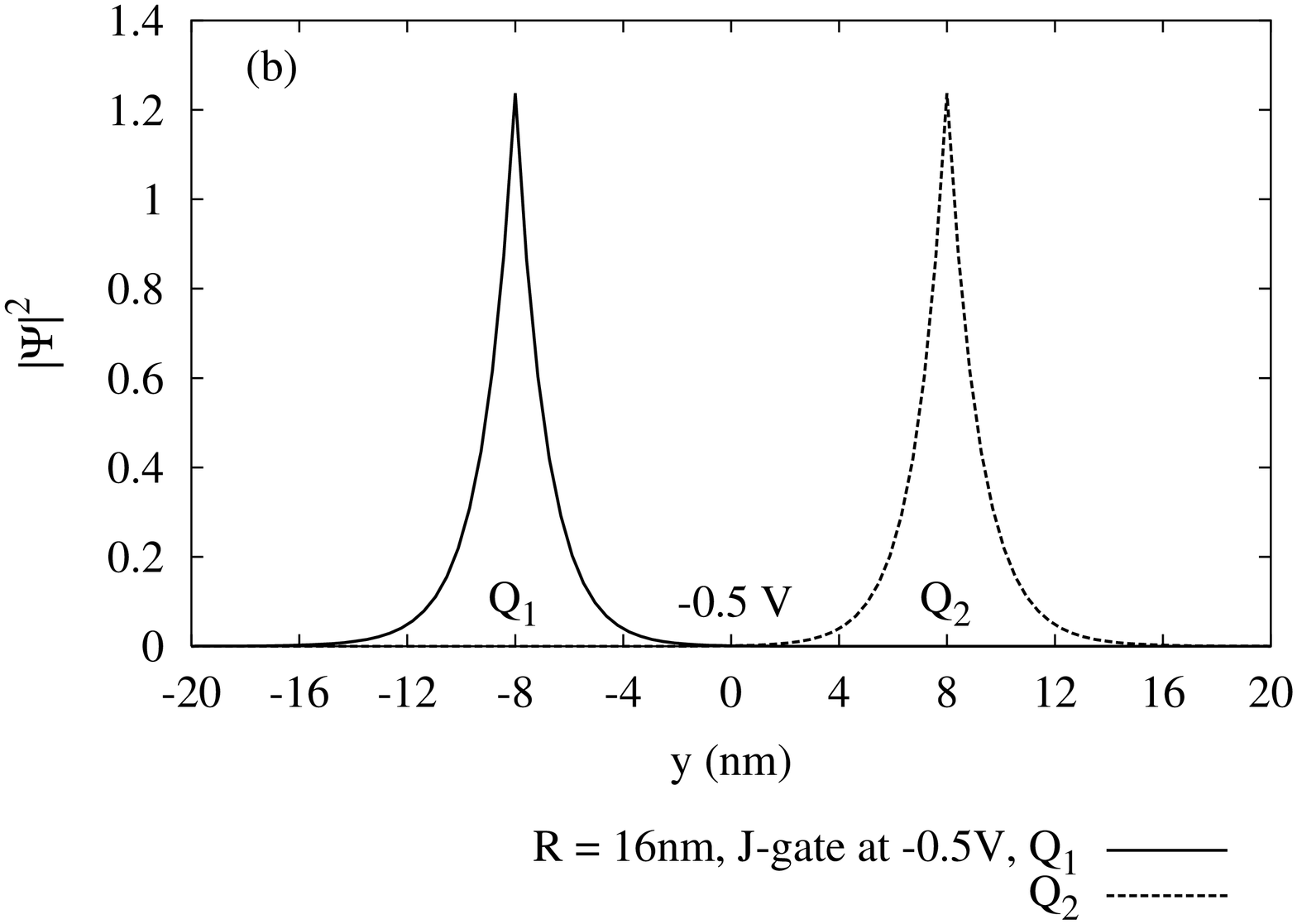} % ,width=7cm,height=11cm,clip=,angle=270}
\end{center}
 \caption{\label{fig:figx6} Ground state electron densities of $Q_1$ and $Q_2$, in $y$-direction at $z=0,x=0$ for an inter donor separation of $R = 14$ and 16nm, a donor depth of $d = 20$nm and a voltage of -0.5 V at the $J$-gate. Note the different vertical scales of $|\Psi|^2$ in (a) and (b). }
\end{figure}

We observe in \fref{fig:figr3} for certain negative gate voltages that the contact hyperfine interaction $A(V)\approx0$, which indicates that the donor wave function has distorted completely away from the nucleus. For $R \leq 14$nm, and $V \leq -0.5$V, the electron is no longer bound to the nucleus, and disperses completely away from the applied voltage. Similarly for $16 \le R \le 20$nm, and $V \leq -0.6$V, the electron is no longer bound to the nucleus.  This is also reflected in \fref{fig:figx6} where we see an abrupt change in the donor wave functions going from $R=14$ to 16nm for a voltage of -0.5 V at the $J$-gate. For $R=14$nm the electron density at the nucleus is close to zero and the wave function disperses from the negative voltage in all directions. In contrast the wave function for $R = 16$nm is still bound to the nucleus and only perturbed slightly by the applied negative voltage. 

The effect of the gate voltage on the donor electron depends on the distance of the qubit from the gate.\cite{me,fran,smit} The donor depth of 20nm and inter donor separations considered in this section ($R\leq20$nm) means that the qubits are situated at relatively short distances from the $J$-gate. So for positive voltages the electron transfer to the gate with increasing $J$-gate voltage is gradual.\cite{smit} However for negative $J$-gate voltages, there is an abrupt change in the electron density at the nucleus for critical negative voltages where the electron is no longer bound to the nucleus.  We find depending on the distance from the gate and the magnitude of the negative gate potential, the electron transfer is either gradual or abrupt.

\subsection{Results for the exchange splitting}

In order to compare our results with previous work\cite{koiller,koiller2,cam,fran,fang,kane2}, we evaluated the zero field exchange interaction, $J(R)$, in bulk Si, for varying inter donor separation.  Our results using effective Bohr radii, $a=2.381$nm and $b=1.368$nm, are in close agreement to the calculations of Fang \emph{et al.}\cite{fang} The zero field exchange splitting calculated here is higher than other reported theoretical values using H-L theory,\cite{koiller,koiller2,cam,fran} because we chose the larger Bohr radius, $a = 2.381$nm, to be along the inter donor axis, and hence the exchange splitting is larger using this convention. The larger Bohr radius was chosen to be along the inter donor axis so that it would also be towards the positive $J$-gate potential, and the smaller Bohr radius in the direction towards the interfaces.

Until calculations are performed which include the effects of the interfaces on the donor wave function, it is hard to verify whether a higher exchange energy would in fact be expected because of the decreased probability of penetration of the donor wave function into the interface regions. Koiller \emph{et al.}\cite{koiller} calculated the exchange coupling in uniaxially strained Si in the presence of interfaces, and also found that these environmental influences could affect the exchange coupling significantly. They found that the $F_\pm (z)$ envelopes were favoured energetically, because the smaller effective Bohr radius in the $z$-direction guarantees less significant penetration of the wave function into the barrier regions. In this paper we are trying to model the effects of the electric field potential and Si host geometry on the donor wave function, to investigate the variation of the exchange splitting with the applied voltage, rather than the absolute values of $J(R)$. 

\begin{figure} [t!]
\begin{center}
 \includegraphics[height=3in,width=2.5in,angle=-90]{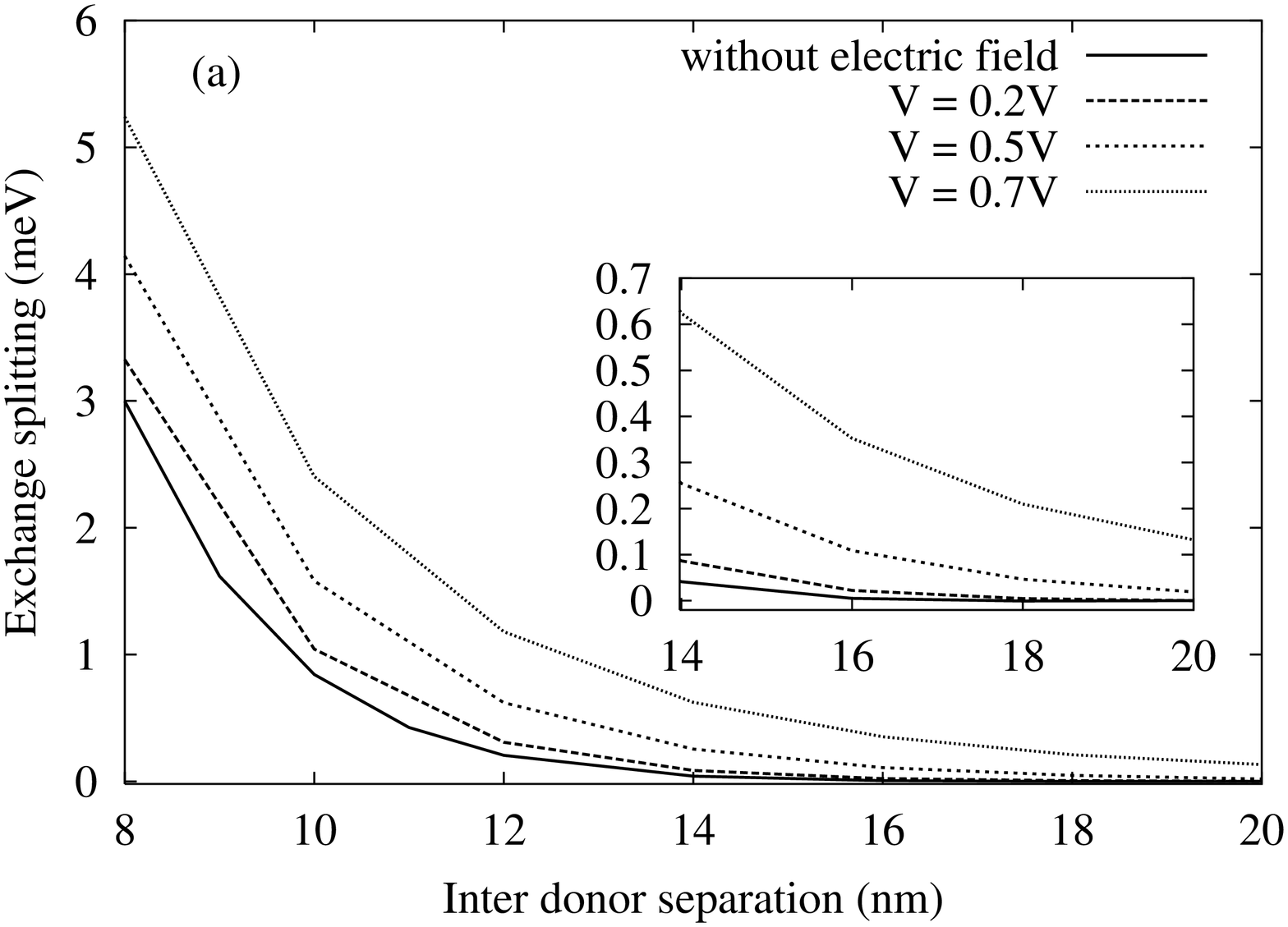} % ,width=7cm,height=11cm,clip=,angle=270}
 \includegraphics[height=3in,width=2.5in,angle=-90]{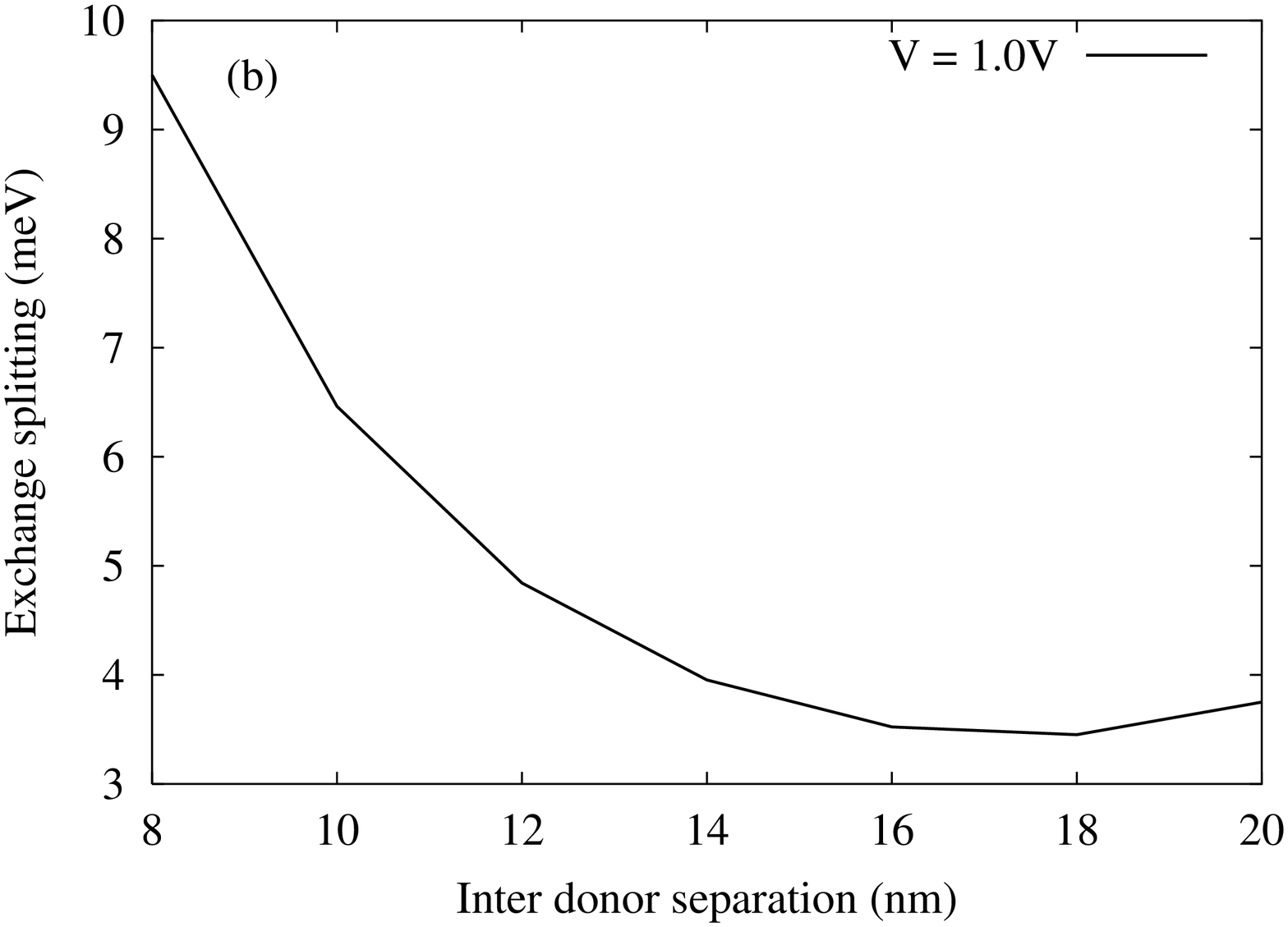} % ,width=7cm,height=11cm,clip=,angle=270}
\end{center}
 \caption{\label{fig:figx2} Calculated exchange coupling as a function of inter donor distance and positive $J$-gate voltage, (a) is the results for $V \leq 0.7$V and (b) for $V = 1.0$V, for a donor depth of 20nm. }
\end{figure}

\Fref{fig:figx2} presents our results for the exchange coupling as a function of inter donor separation and positive $J$-gate voltage. We observe that the exchange coupling increases as the $J$-gate voltage increases as expected, since the applied field draws the electrons closer together. At a voltage of 1.0V the donor electron wave function is perturbed the greatest, and the exchange coupling is significantly higher at this voltage for every inter donor separation. 

At all voltages lower than 1.0V, the exchange coupling decreases as $R$ increases as expected, but  in \fref{fig:figx2}(b) for a voltage of 1.0V and for  $R =20$nm,  the exchange coupling actually increases slightly. This is because at large inter donor separations the donor is further from the $J$-gate, and thus is more attracted to the potential well at the $J$-gate as the gate voltage increases sufficiently. Hence the overlap between the adjacent donor electron orbitals is slightly greater, even if the inter donor separation is higher.

\begin{figure} [t!]
\begin{center}
 \includegraphics[height=3in,width=2.5in,angle=-90]{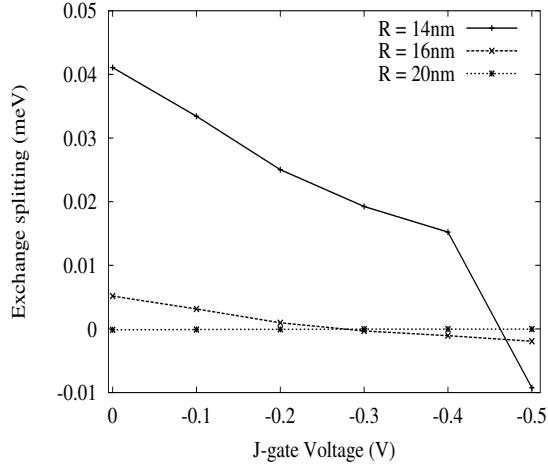} % ,width=7cm,height=11cm,clip=,angle=270}
\end{center}
 \caption{\label{fig:figx3} Calculated exchange coupling as a function of inter donor distance and negative $J$-gate voltage, for a donor depth of 20nm. }
\end{figure}

 One of the advantages of the Kane quantum computer is the ability to turn on or off the coupling between the different qubits. \Fref{fig:figx3} presents our results for the exchange coupling as a function of inter donor separation and negative $J$-gate voltage. The exchange coupling decreases as the negative applied potential decreases. When the applied negative voltage is large enough, the electron is greatly distorted away from the nucleus. We have seen this in \fref{fig:figx6}(a) that at a $J$-gate voltage of -0.5V and $R=14$nm, the donor wave functions for $Q_1$ and $Q_2$ have perturbed away from the applied voltage in opposite directions. In this case we have effectively turned off the coupling between the adjacent qubits, as the overlap between the two electron densities is almost zero.  
 
Fang \emph{et al.}\cite{fang} used a spherical effective mass Hamiltonian and modeled the  $J$-gate potential using a one dimensional parabolic well, and in their work they do not consider the effect of interfaces. In their calculations of the exchange splitting they considered relatively large inter donor separations ($\geq$ 16nm) and a smaller electric field potential at the $J$-gate relative to the TCAD cross-sectional potential we obtain between the donors. This made it hard to compare to our work, but as a rough estimate we can compare our results for a voltage of 0.2 V at the $J$-gate, where the TCAD potential at the mid-point between the two donors, approximately 0.02 V, is nearly equal to the potential at the minimum well used by Fang \emph{et al.}\cite{fang} for $\mu = 0.6$. 

We observe that our calculations are up to an order of magnitude lower than those calculated by Fang \emph{et al.} This discrepancy is probably because the electric field potentials used by our method and Fang \emph{et al.} are not equivalent, as we obtain the electric field potential within the whole device, whereas they use a simplified 1-D potential. We also compared our results with Parisoli \emph{et al.}\cite{fran,cam} who used a spherical effective mass Hamiltonian, with effective mass, $m_* \approx 0.29m_0$ and effective Bohr radius, $a \approx 2$nm. The results for the exchange coupling agreed qualitatively (to within an order of magnitude) and predicted the same trend in variation of the exchange coupling with voltage. Again our results were consistently higher because of the larger effective Bohr radius along the inter donor axis.

The application of a voltage to the $J$-gate, and the magnitude of the qubit separation, can be used to control the strength of the exchange coupling of the donor pair. At lower voltages the most significant factor influencing the exchange coupling is the inter donor separation. As the positive voltage is increased, the donor electron wave function is perturbed greater as the donor is moved further away from the gate, and the exchange coupling can be enhanced even with large qubit separations. At smaller inter donor separations the donor electron is more affected by negative voltages. For voltages lower than a certain critical value, the donor electron is completely transferred away from the nucleus and the negative voltage, and the exchange coupling decreases almost to zero.

\section{\label{sec:fivea} Results obtained varying the donor depth}

We observed the effect that the donor depth, $d$, below the silicon oxide layer has on the donor ground state perturbed by a gate voltage. Here we ran calculations for $5 \leq d \leq 45$nm.

\subsection{Results for the contact hyperfine interaction}

\begin{figure}
\begin{center}
 \includegraphics[height=3in,width=2.5in,angle=-90]{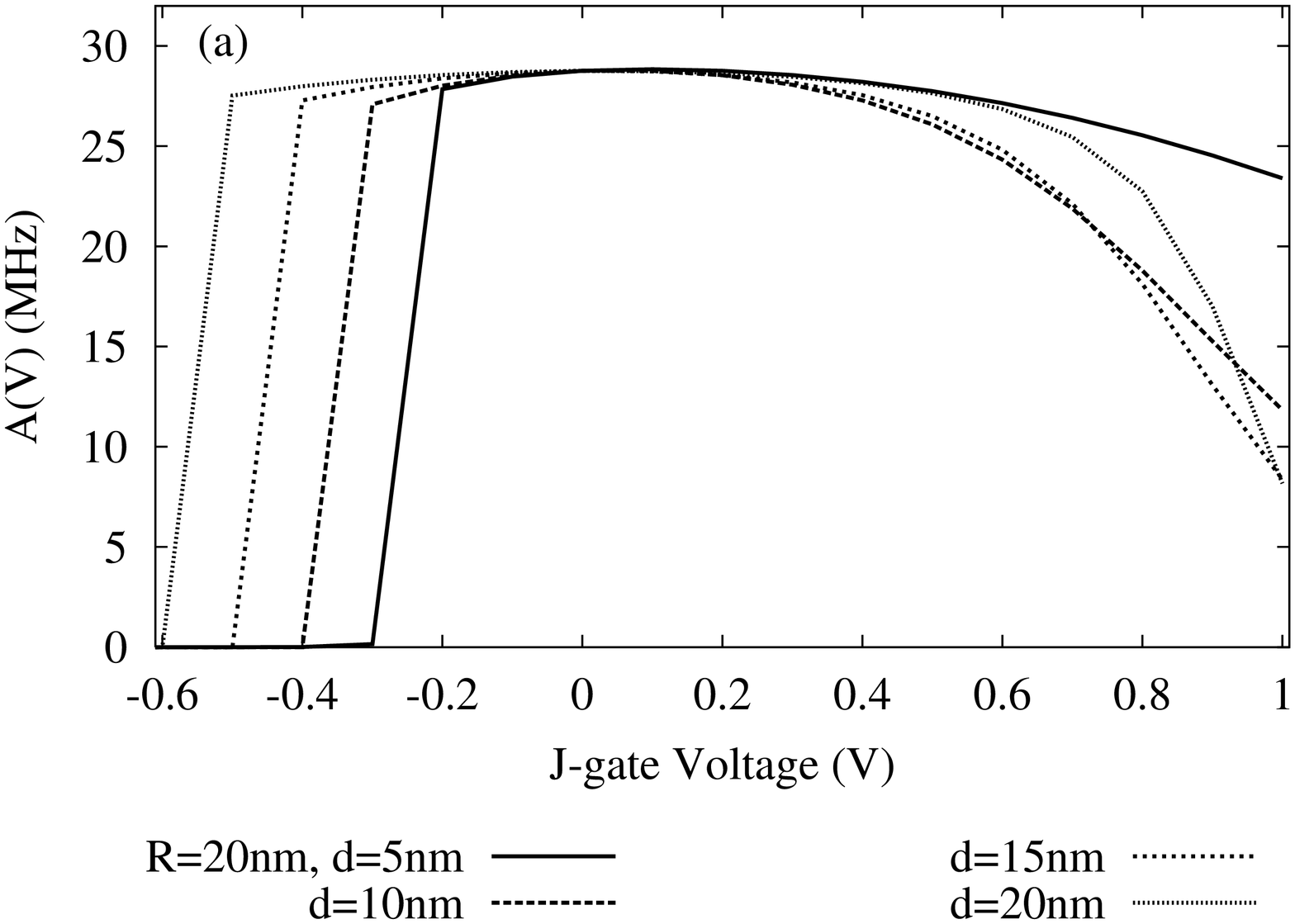} % ,width=7cm,height=11cm,clip=,angle=270}
\includegraphics[height=3in,width=2.5in,angle=-90]{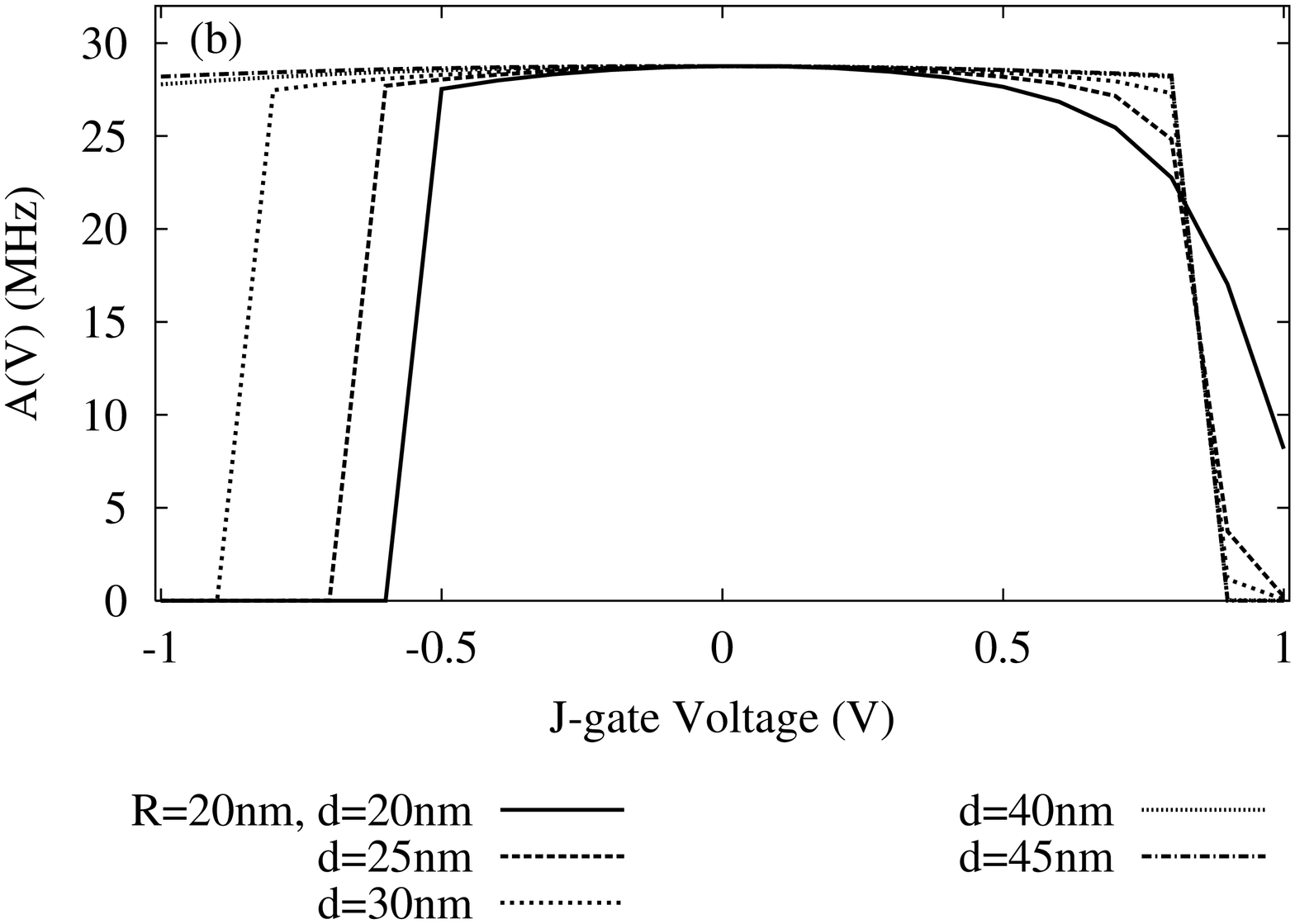} % ,width=7cm,height=11cm,clip=,angle=270}
\end{center}
 \caption{\label{fig:figv4}   Contact hyperfine interaction at varying donor depths and $J$-gate voltage, with inter donor separation $R =$ 20nm. }
\end{figure}

\Fref{fig:figv4} show our results for the contact hyperfine interaction at varying donor depth and gate voltage. Here we observe similar trends in the variation of the hyperfine interaction as in the previous section. We see that as $d$ increases, and hence the distance from the $J$-gate also increases, we see a cross-over behaviour where the donor wave function is perturbed greater for larger donor depths at positive voltages above a critical value. Also for $d \geq 25$nm we see in \fref{fig:figv4}(b) there is an abrupt decrease in the electron density, defining an ionisation voltage at these donor depths. This process of ionisation has been reported previously.\cite{me, smit}

\begin{figure} [t!]
\begin{center}
 \includegraphics[height=3in,width=2.5in,angle=-90]{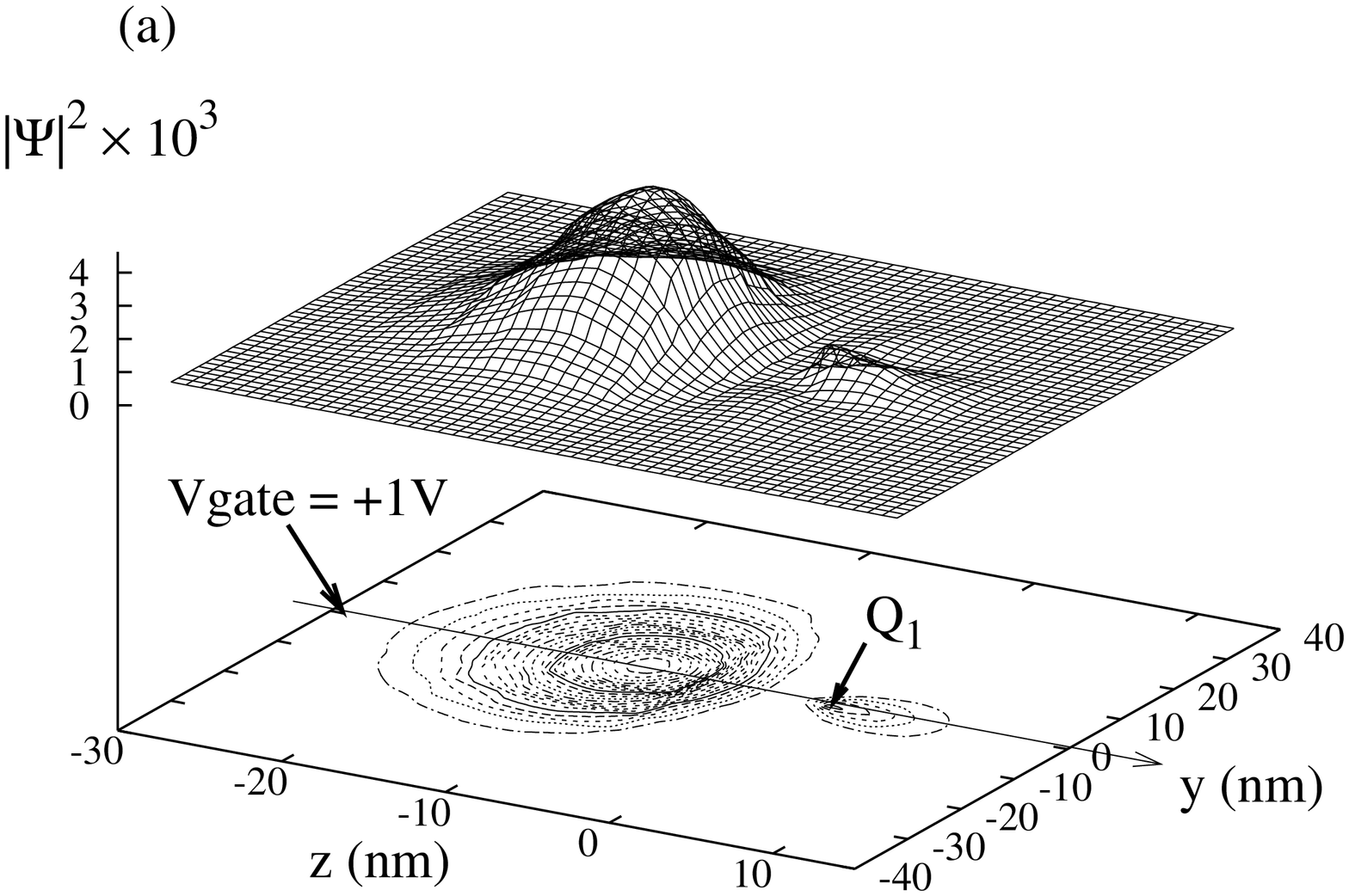} % ,width=7cm,height=11cm,clip=,angle=270}
 \includegraphics[height=3in,width=2.5in,angle=-90]{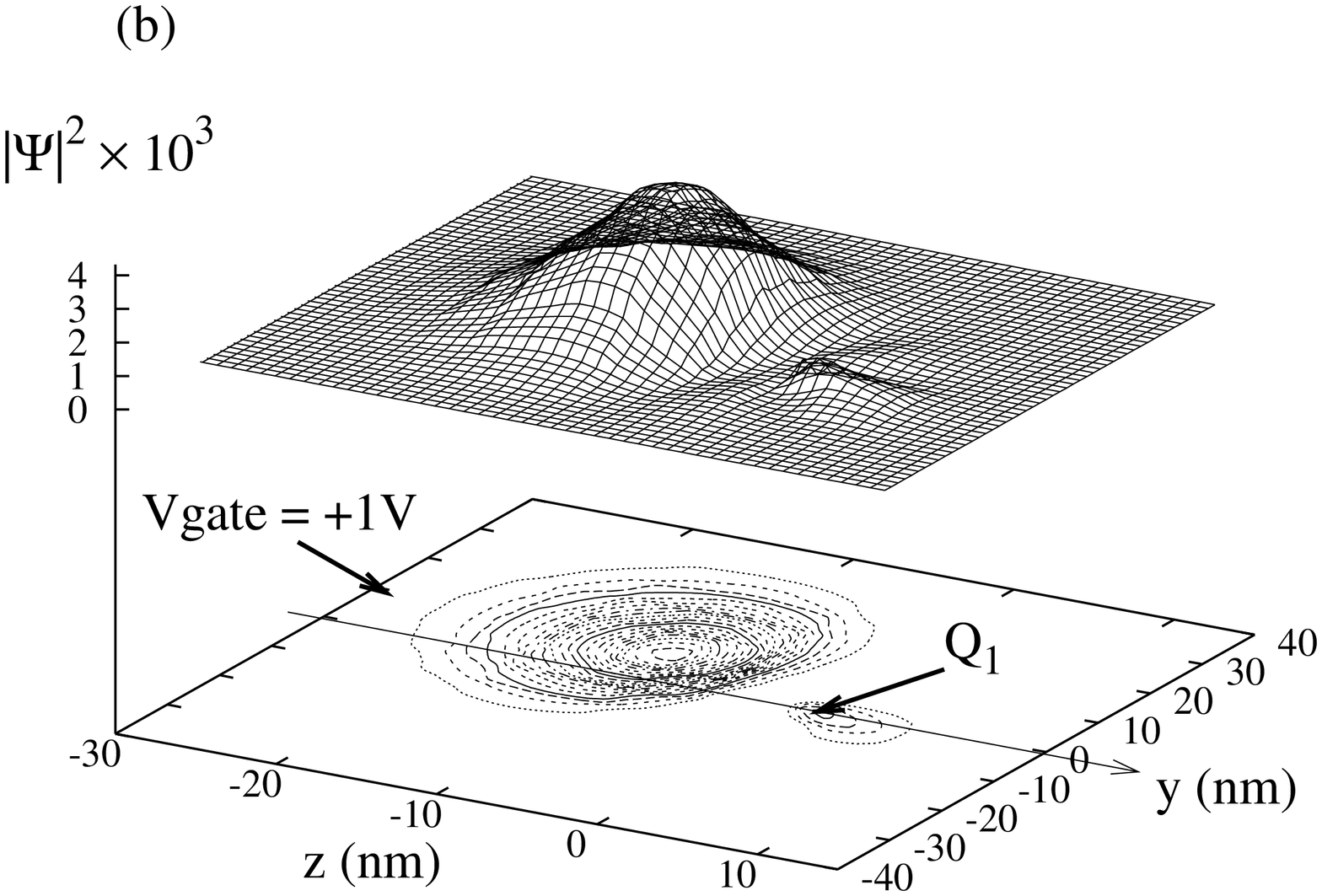} % ,width=7cm,height=11cm,clip=,angle=270}
\end{center}
 \caption{\label{fig:figv2}  Ground state electron density in $yz$-plane for donor depth at 30nm with 1.0V at the $A$-gate in (a) and $J$-gate with $R=20$nm for (b). In both plots $Q_1$ is located at the origin, and we have included the $y=0$ symmetry line in the contour plot, to highlight the difference in the electron density for an applied $A$ or $J$-gate voltage.}
\end{figure}

For $d \geq 25$nm and large enough positive voltages, the electron has perturbed or ionised completely to the gate. Figures~\ref{fig:figx4} and~\ref{fig:figv2} show the contrast in the donor wave function for different donor depths. In \fref{fig:figv2} at $d=30$nm, the electron is perturbed almost completely away from the nucleus towards the applied voltage. While in \fref{fig:figx4} at $d=20$nm, the donor wave function is only slightly perturbed from the zero field ground state by the applied voltage. 

The basis we are using for the donor electron wave function consists only of bound states, which is a good approximation for the smaller gate voltages, as the electron is still bound to the nucleus. To model the ionisation process at larger gate voltages more accurately, a more rigorous approach would be to include the delocalised conduction band states in the basis as well. 

\Fref{fig:figv2}(a) and (b) shows the comparison between the electron ground state probability density in the $yz$-plane, obtained for a voltage of 1.0V applied to the $A$ and $J$-gate respectively, for a donor depth of 30nm and inter donor separation of $R=$20nm. For an $A$-gate voltage the donor wave function is symmetric in $y$ and only perturbs toward the $A$-gate in the $z$ direction. In comparison, when a $J$-gate voltage is applied, the wave function can distort in both the $y$ and $z$-directions.  Unfortunately as the donor depth becomes greater, selectivity may be lost, and a voltage applied at either the $A$ or $J$-gate will cause the same change in the contact hyperfine interaction.

\subsection{Results for the exchange splitting}

\begin{figure}
\begin{center}
\includegraphics[height=3in,width=2.5in,angle=-90]{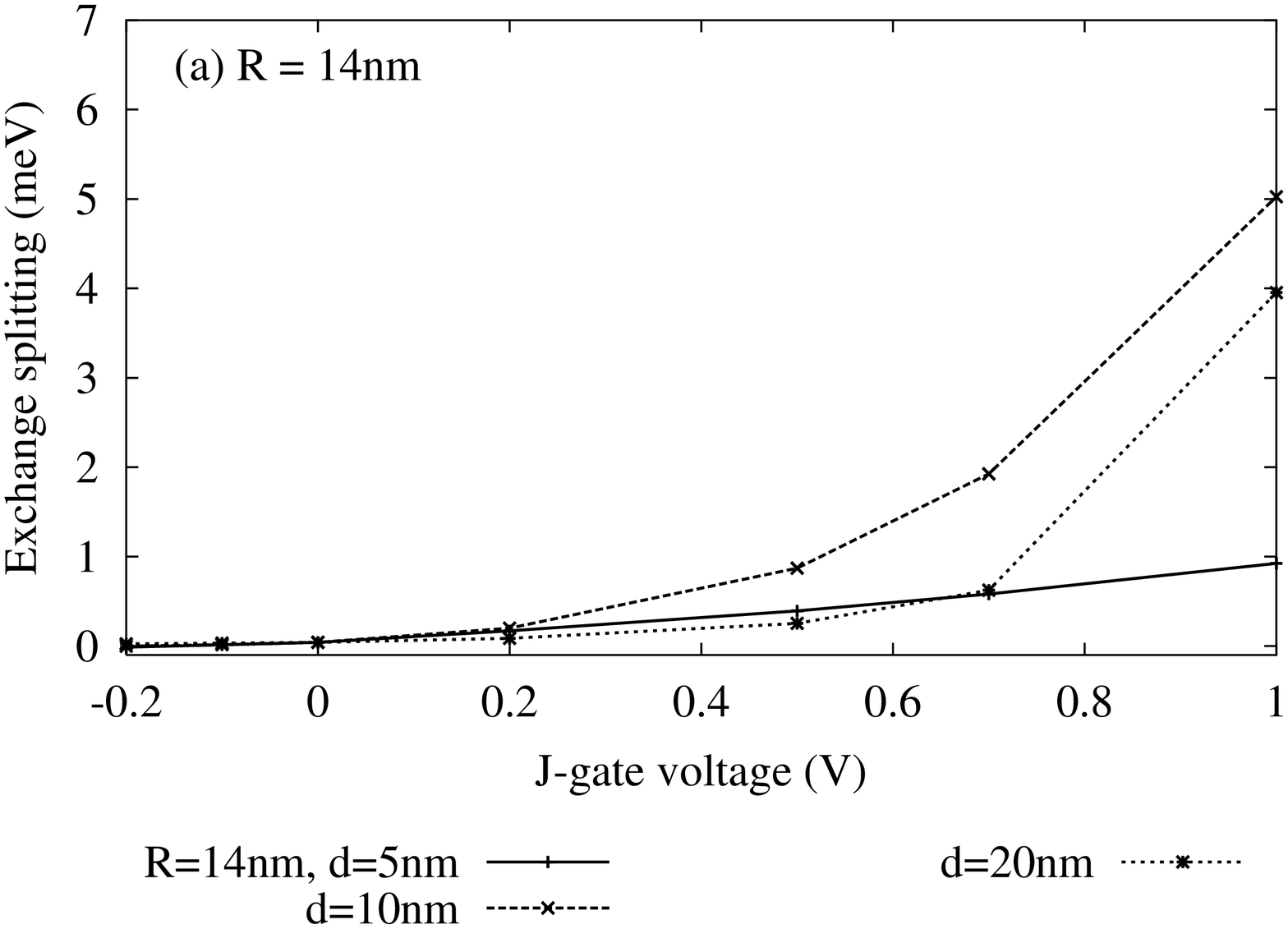} % ,width=7cm,height=11cm,clip=,angle=270}
\includegraphics[height=3in,width=2.5in,angle=-90]{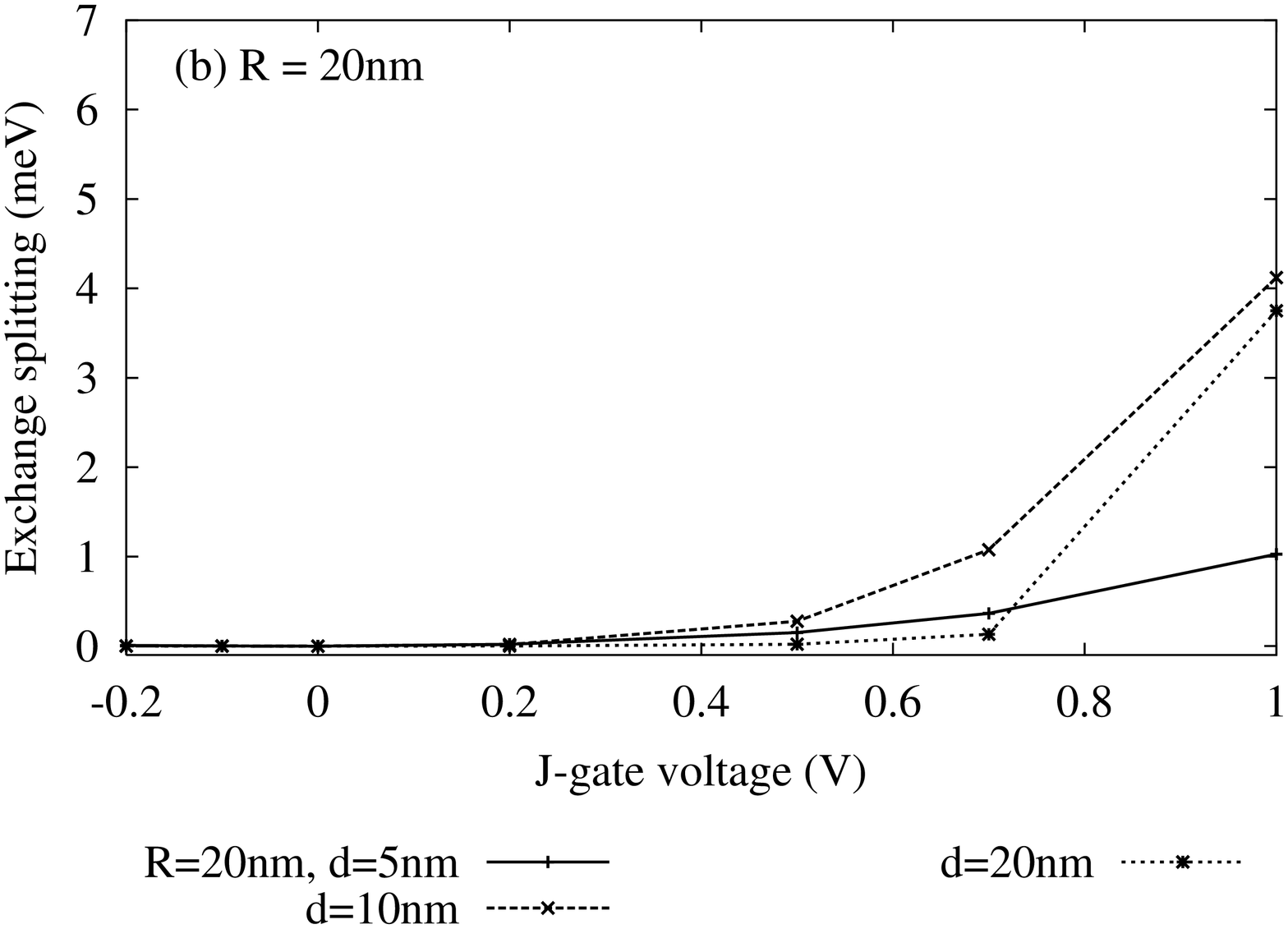} % ,width=7cm,height=11cm,clip=,angle=270}
\end{center}
 \caption{\label{fig:figx5} Calculated exchange coupling as a function of donor depth and $J$-gate voltage, with $R$=14nm in (a), and $R$=20nm in (b).}
\end{figure}

\Fref{fig:figx5} shows the variation of the exchange coupling with donor depth and voltage for two inter donor separations, $R=14$ and 20nm. It is evident that the depth of the donor influences the degree to which the electron is perturbed by the gate voltage, and hence will also affect the strength of the exchange coupling. 

For small $d$ the electron is only slightly perturbed by the positive gate voltage, as the P nucleus and the gate voltage are so ``strongly coupled"\cite{smit} that the effect of the gate voltage is only to further stabilise the donor electron. So we see for $d=5$nm that although the exchange coupling has increased significantly from the zero field coupling, it is still not as strong as the coupling for $d=10$ and 20nm. 

The exchange coupling for $d=10$ and 20nm are similar for equal inter donor separations, and the effect of the magnitude of $d$ is not so pronounced. Here we can see that for a donor depth of 10nm the exchange coupling is enhanced the most by the applied voltage. One of the reasons for this may be that for $d=10$nm the donor wave functions predominantly move toward the applied voltage at the $J$-gate in the $y$-direction along the inter donor axis and thus the exchange coupling is enhanced further, whereas for $d=20$nm the wave functions can perturb in both the $y$ and $z$-directions toward the $J$-gate.

\section{\label{sec:six}Conclusions and prospects for achieving silicon-based quantum computation}

In this work we have studied the P donor wave function perturbed by an electric field and the Si host geometry, and the two interactions fundamental to the Kane quantum computer: the hyperfine and exchange interactions. We have studied the effect of varying several experimental parameters: the gate voltage, inter donor separation, and donor depth in order to fine tune the hyperfine and exchange interactions.

The results presented highlight the significance of not only the gate potential in affecting the donor electron wave function, but also the position of the qubits in the device. One of the critical discoveries was that the inter donor separation is not the only relevant factor in determining the strength of the exchange coupling, the proximity of the qubit to the gate is also important in determining the degree to which the electron exchange interaction can be enhanced by the applied voltage. 

In the absence of an electric field, only the inter donor separation is instrumental in determining the strength of the exchange coupling, and as $R$ increases the exchange coupling decreases. However, when a large positive voltage is applied at the  $J$-gate, either a gradual transference of the donor electron density occurs for dopants close to the gate, and the exchange coupling is enhanced proportionally. Or if the electron is ionised by the gate voltage the exchange coupling can be enhanced considerably even for quite large inter donor separations and donor depths. So both of these competing influences must be considered in modeling the strength of the exchange coupling.

For the parameters we studied (ie. $R \leq 20$nm) and negative $J$-gate voltages, $V \leq -0.6$V, the electrons at $Q_1$ and $Q_2$ disperse away from the applied negative voltage at the $J$-gate in opposite directions. In this case the overlap between the two electron densities is almost zero, and the exchange coupling also decreases almost to zero, and we have effectively turned off the coupling between adjacent qubits.

Future developments in our laboratory is concentrating on confirming these results using a more rigorous evaluation of the exchange coupling. However this initial work provides valuable insight into the environmental influences which play a vital role in determining the sensitivity of the donor electron wave function to the applied electric field and hence the sensitivity also of the exchange coupling and the hyperfine interaction. 

\ack
L M Kettle and H-S Goan would like to thank C I Pakes and G J Milburn for valuable discussions relating to this work. This work was supported in part by the Australian Partnership for Advanced Computing National Facility, the Australian Research Council, the Australian Government, the US National Security Agency, the Advanced Research and Development Activity, and the US Army Research Office under contract number DAAD19-01-1-0653. H-S Goan would like to acknowledge support from a Hewlett-Packard Fellowship.

\end{document}